\documentclass [floatfix, superscriptaddress, reprint, showpacs, aip, apl, 10pt]{revtex4-1}

\usepackage{amsmath}
\usepackage{hyperref}
\usepackage{graphicx}
\usepackage{bbm}
\usepackage{amssymb}
\usepackage[english]{babel}
\usepackage[latin1]{inputenc}
\usepackage{lmodern}
\usepackage[T1]{fontenc}
\usepackage{xcolor}
\usepackage{mathtools}

\graphicspath{{Figs/}}

\begin{document}

\title{Self-adaptive loop for external disturbance reduction in differential measurement set-up}

\author{Giuseppe Bevilacqua}
\author{Valerio Biancalana} 
\email{valerio.biancalana@unisi.it}
\affiliation{DIISM, University of Siena, Via Roma, 56 - 53100 Siena (Italy)}

\author{Yordanka Dancheva}
\author{Antonio Vigilante}
\affiliation{DSFTA, University of Siena, Via Roma, 56 - 53100 Siena (Italy)}

\begin{abstract}
We present a method  developed to actively compensate common-mode magnetic disturbances on a multi-sensor device devoted to differential measurements. The system uses a field-programmable-gated-array card, and operates in conjunction with a high sensitivity magnetometer: compensating the common-mode of magnetic disturbances results in a relevant reduction of the difference-mode  noise. The digital nature of the compensation system allows for using a numerical approach aimed at automatically adapting the feedback loop filter response. A common mode disturbance attenuation exceeding  50 dB is achieved, resulting in a final improvement of the differential noise floor by a factor of 10 over the whole spectral interval of interest.

\end{abstract}

\date{\today}

\maketitle

\section{Introduction}

	Magnetic field variations often constitute a relevant disturbance in many fields of research among which: of an applied nature - measurement of biomedical signals  \cite{johnson_apl_10, malmivuo_jpe_87, boto_ni_17}, ultra-low-field (ULF) NMR spectroscopy \cite{sjolander_jpcl_17,  tayler_rsi_17},  magnetic resonance imaging \cite{savukov_apl_13}, geophysics, archaeology, security \cite{cooper_prappl_16, deans_prl_18}, etc., and of a fundamental nature - the search of dark matter and dark energy constituents \cite{kimball_prd_18}, determination of neutron electric dipole moment (nEDM) \cite{pendlebury_prd_15}, etc.	
	
	Different physical parameters of the system have to be considered when counteracting magnetic disturbances: according to the particular application, given specifications must be taken into account (as for example attenuation factor, bandwidth, long term stability, geometric constraints, weight and size, energy consumption etc.). 
	
	Various methodologies for magnetic disturbance reduction have been developed, based on diverse approaches. Such methodologies can be applied in other fields, where the achieved sensitivity is affected by the external noise sources as in the case of electronic or acoustic measurements. 
	
	These methodologies can be roughly categorized in three families: shielding, compensation, and cancellation by difference \cite{Jiang_2018J, gerginov_josab_17, cooper_jmr_18}. In the first two cases the external noise is reduced before the measurement, while in the third one it is measured and subsequently subtracted. The shielding approach makes use of passive devices stopping/attenuating the disturbance, while compensation approach is generally based on feedback techniques, and thus requires a secondary sensor driving a system that actively reacts to the disturbance. 
	
	The three mentioned methodologies are not mutually exclusive. Indeed, excellent results have been obtained  by opportunely combining systems based on different approaches. As an example, in the nEDM research the stability and homogeneity of the magnetic field inside the magnetic shield depends, to a large extent, on the magnetization state of the high-permittivity material constituting the passive shield, which is in turn affected by the external noise. Thus an important improvement  is achieved by actively stabilizing the external field \cite{afach_jap_14}. 
	
	Several kinds of sensors have been used for extracting the error signal in case of magnetic field compensation. Their selection is of primary importance because it determines and limits the precision, the accuracy and the bandwidth of the compensation system. The literature reports cases of pick-up sensors based on superconducting quantum interference devices (SQUID) \cite{brake_mst_1993}, fluxgates \cite{deans_rsi_18}, atomic sensors \cite{belfi_rsi_10}, and magnetoresistances \cite{ringot_pra_01}.  
	
	As an actuator, most of the compensation systems described in the literature make use of coils. The sensor signal is usually amplified and fed back to the coils through a compensation loop containing a proportional or a proportional-integral-derivative control. The poor knowledge of the frequency response of the sensor, the actuator and the electronics used to close the loop prevents the \textit{a-priori} design of accurate loop filters, so that some tuning is eventually necessary, that is commonly performed following empirical procedures, as --for example-- in the case of Ref. \cite{afach_jap_14}. 
	
	In this work we propose a novel feedback control system implemented in a field programmable gate array (FPGA) that numerically filters the error signal, so to enable rearrangement and fine tuning of the loop filter. This feature is profitably automated by means of a procedure that adapts the loop response to optimize its performance. The resulting self-adaptive feedback contains a filter network where the number, the arrangement, the types and the characteristic frequencies of the filters can be adjusted.

	In other terms, the optimization procedure works on a multi-parameter loop, and finds the best tuning and topology of the loop filter, adapting the feedback to the responses of the sensor and of the actuator.

    The proposed self-adaptive active compensation system is developed to upgrade a multi-channel optical atomic magnetometer (OAM) \cite{bevilacqua_apb_16}, used to perform differential (gradiometric) measurements. The whole apparatus is designed to operate in an unshielded volume, and is devoted to detect weak signals from the difference mode (DM) response, while using the common mode (CM) term as an error signal.

Most of the OAMs with ultra-high sensitivity are operated in shielded volumes, not only to counteract the environmental noise, but also to operate at nT levels, which are necessary to run the most sensitive implementations  \cite{ledbetter_08}. However, there are also applications for which unshielded operation is required. The development of OAMs operating in unshielded (even if compensated) environment is of strategic importance whenever highly sensitive portable detectors, are used in areas like security, magnetometric characterizations of samples that are parts belonging to bulky objects.

\section{Setup}
\subsection{The magnetometer}

Among the big variety of magnetic disturbances we are interested on those originating both from human and natural activities in the frequency range from 0.1 to a few hundred Hz. In this work a multi-channel OAM sensor is profitably coupled with a magnetic field stabilization system, where the error signal is provided by one of the channels. The magnetometer operates by measuring the magnetic field difference between two channels (DM mode). The analyzed sample is placed next to one of them, thus it affects mainly the DM signal, while disturbances, from far located sources, give mainly CM terms. 

Compensating the CM term in DM measurements brings a two-fold advantage: the limited CM rejection ratio makes residual CM disturbance appear as a noise affecting also the DM signal; moreover in some applications  the CM drives the dynamics of the analyzed sample: for example, in the ULF NMR experiments \cite{bevilacqua_jpcl_17} the signal originates from the nuclear precession, which is directly affected by the CM disturbance.

\begin{figure}[ht]
   \centering
    \includegraphics [angle=270, width= 0.8 \columnwidth] {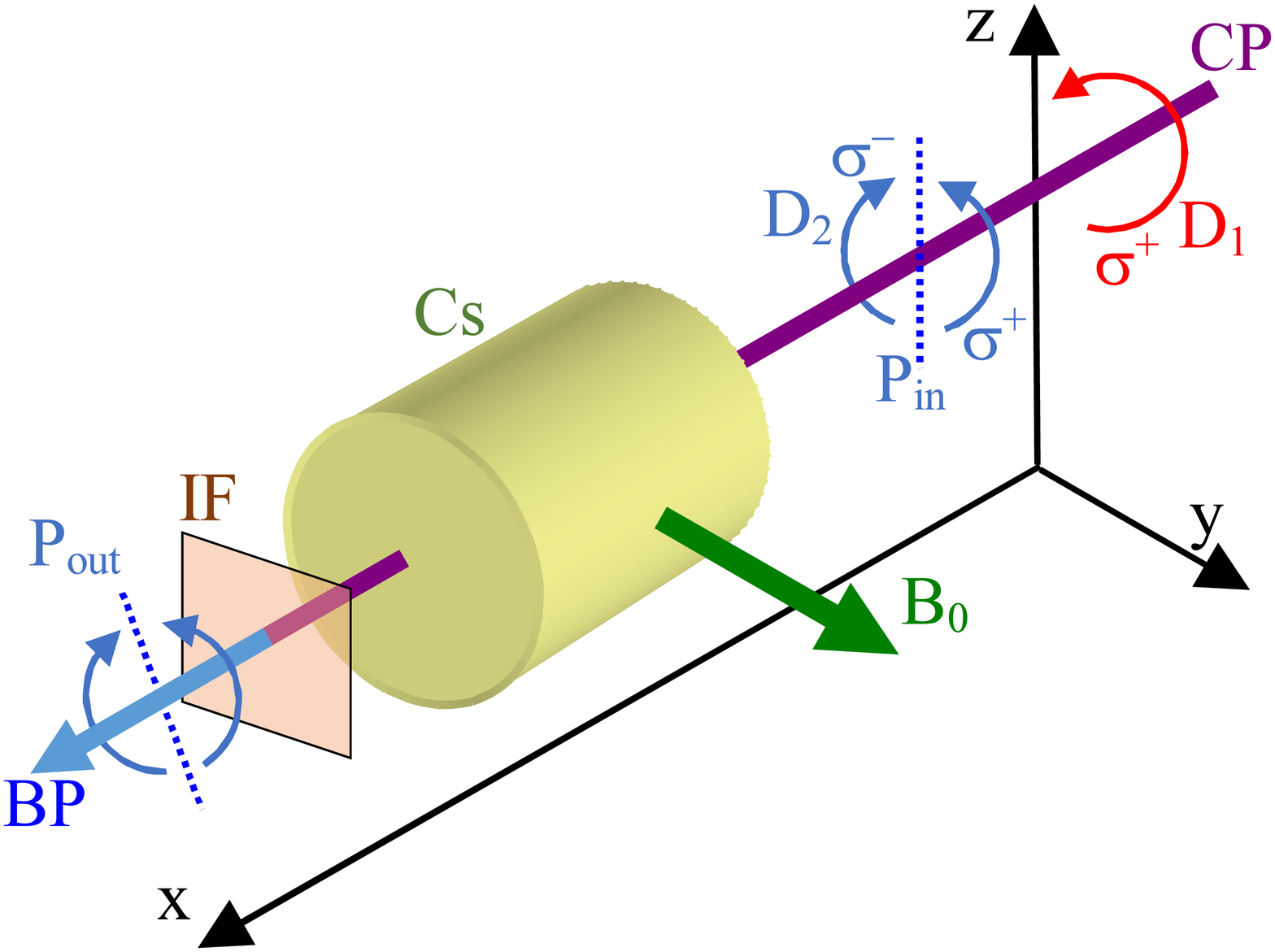}
  \caption{Single channel OAM schematics. The Cs cell is illuminated by two types of co-propagating radiation, which are appropriately polarized to pump (red curved arrow) and probe (blue curved arrows) the atomic magnetization. The pump and probe beams denoted as CP are co-propagating through the Cs cell. An interference filter IF is used to block the pump beam. The Faraday rotation of the probe's polarization (from P$_{\mathrm{in}}$ to P$_{\mathrm{out}}$) is measured by a balanced polarimeter BP.
  \label{fig:setup}}
\end{figure}

The OAM considered in this work \cite{bevilacqua_apb_16} uses vapor cells with a solid source of Cs and 23~Torr of $N_2$ buffer gas that are warmed up to about~ 45\textdegree{C}.  The magnetometer operates in a homogeneous bias magnetic field $\vec B_0$ (around one $\mu$T and below)  obtained through a set of large (180~cm size) Helmholtz coils. The set, composed by three orthogonal coils pair, partially compensates the static environmental field. A bias field along \textbf{y} axis (see Fig.\ref{fig:setup}) remains. The static field inhomogeneities are also zeroed through five electromagnetic quadrupoles \cite{biancalana_rsi_17}.

The Cs atoms are optically pumped into a stretched (maximally oriented) state by means of laser radiation at a milli-Watt level. This pump radiation is circularly polarized (red curved-arrow in Fig.\ref{fig:setup}) and tuned to the D$_1$ line. After the interaction with the Cs atoms, the pump beam is stopped by an interference filter IF. The time evolution of the atomic state is probed by a weak (micro-Watt level), linearly polarized light (blue curved arrows in Fig.\ref{fig:setup}) tuned to the proximity of the D$_2$ line. The pump and probe beams co-propagate along the \textbf{x} direction (see Fig.\ref{fig:setup}).

The transverse magnetic field $\vec B_0$ causes a Larmor precession of the induced magnetization at $\omega_L=\gamma B_0$ (where $\gamma$ being gyromagnetic factor). The magnetization decay is counteracted through synchronous optical pumping, that is obtained by modulating the pump laser wavelength at a frequency $\omega_M/2\pi$ ($\omega_M$ is resonant with $\omega_L$).

The precession causes a time-dependent Faraday rotation of the probe polarization plane. That rotation is driven to oscillate at $\omega_M$ (forcing term), to which it responds with a phase $\varphi (t)$ evolving in accordance with the magnetic field.

The Faraday rotation in each channel is measured by a balanced polarimeter, which analyzes the probe beam polarization, splitting it in two cross-polarized beamlets. The split beamlets impinge on two photo-diodes, whose photo-current unbalance is a measure of the Faraday rotation.

The system is equipped with four channels, made out of two vapour cells, thus enabling differential measurements over two different baselines: $5.6$~cm and $0.5$~cm. 

\subsection{Error signal extraction and manipulation}

Under operating conditions, the transverse magnetization relaxation rate is inferred from the resonance width and amounts to $\Gamma \approx 2\pi\times 25$~rad/sec. The polarimetric signal of one channel is modeled as: 
\begin{equation}
  A \exp {i(\omega_Mt+\varphi(t))}. 
  \label{eq:signal}
\end{equation}

\begin{figure}[htbp]
   \centering
  \includegraphics [width=   \columnwidth] {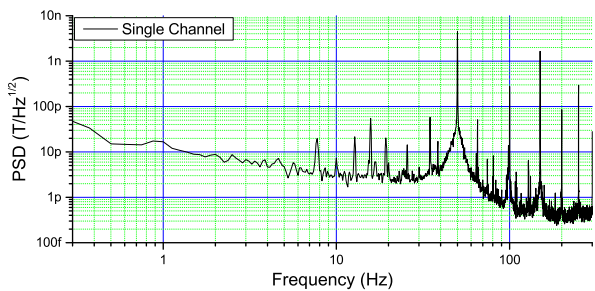}
  \caption{Single channel noise power spectral density (PSD). 
  }
  \label{fig:CMPSD}
\end{figure}

The measured polarimetric signal is real-time numerically demodulated  multiplying it by a reference sinusoid at $\omega_M$, synchronous with laser modulation signal. The resulting $2\omega_M$ terms are notched out by summing the data of an integer number $N$ of periods. A delay of $\Delta_1= N \pi/\omega_M$ 
in the phase determination is consequently introduced. The phase $\varphi$(t) in Eq.(\ref{eq:signal}) is thus extracted, and in a first-order approximation it is expected to respond to a time-dependent field variation $\delta B_{\parallel}(t)$ according to \cite{bevilacqua_apb_16}:

\begin{equation}
   \varphi(t) =\frac{1}{\Gamma}\Big(\omega_L-\omega_M+\omega_1(t)+\dot{\varphi}\Big),
   \label{eq:phase}
\end{equation}

\noindent where $\omega_1=\gamma \;  \delta B_{\parallel}(t)$ is set by the disturbance component $\delta B_{\parallel}(t)$ parallel to  $\vec B_0$. After its conversion into magnetic units, the registered  spectrum appears as shown in Fig.\ref{fig:CMPSD}.

In the typical (unshielded) operating conditions, when signals from multiple channels are analyzed, the CM term amply exceeds the DM terms, so that the noise spectrum of all the channels (and hence the CM) substantially  matches that presented in Fig.\ref{fig:CMPSD}.

\subsection{Feedback loop design}

The FPGA module used in this work, 
 is a commercial device (NI 9147), equipped with
a 16~bit 500~KSa/s analog-to-digital converter (ADC) (NI9222) and two 16 bit 100~KSa/s digital-to-analog converters (DAC)  (NI9263). It is programmed to: demodulate the signal of Eq.(\ref{eq:signal}); determine the phase $\varphi$ according to Eq.(\ref{eq:phase}); convert it to magnetic field (that constitutes the error signal); and filter the error signal to drive an analogue current amplifier supplying the compensation coils. The low-latency digital signal processing \cite{leibrandt_rsi_15} offered by modern electronics makes the limited bandwidth of the digital approach a secondary (but non-negligible) problem at the Larmor frequencies in question (1-20~kHz). It is crucial for the FPGA to execute ADC, data processing, and DAC operation as quickly as possible, because the cycle duration sets an additional delay time in the feedback loop. In our implementation, a cycle duration as short as  $\Delta_2=14\,\mu$s is achieved.

\begin{figure}[htbp]
  \centering
   \includegraphics [angle=270, width=   \columnwidth] {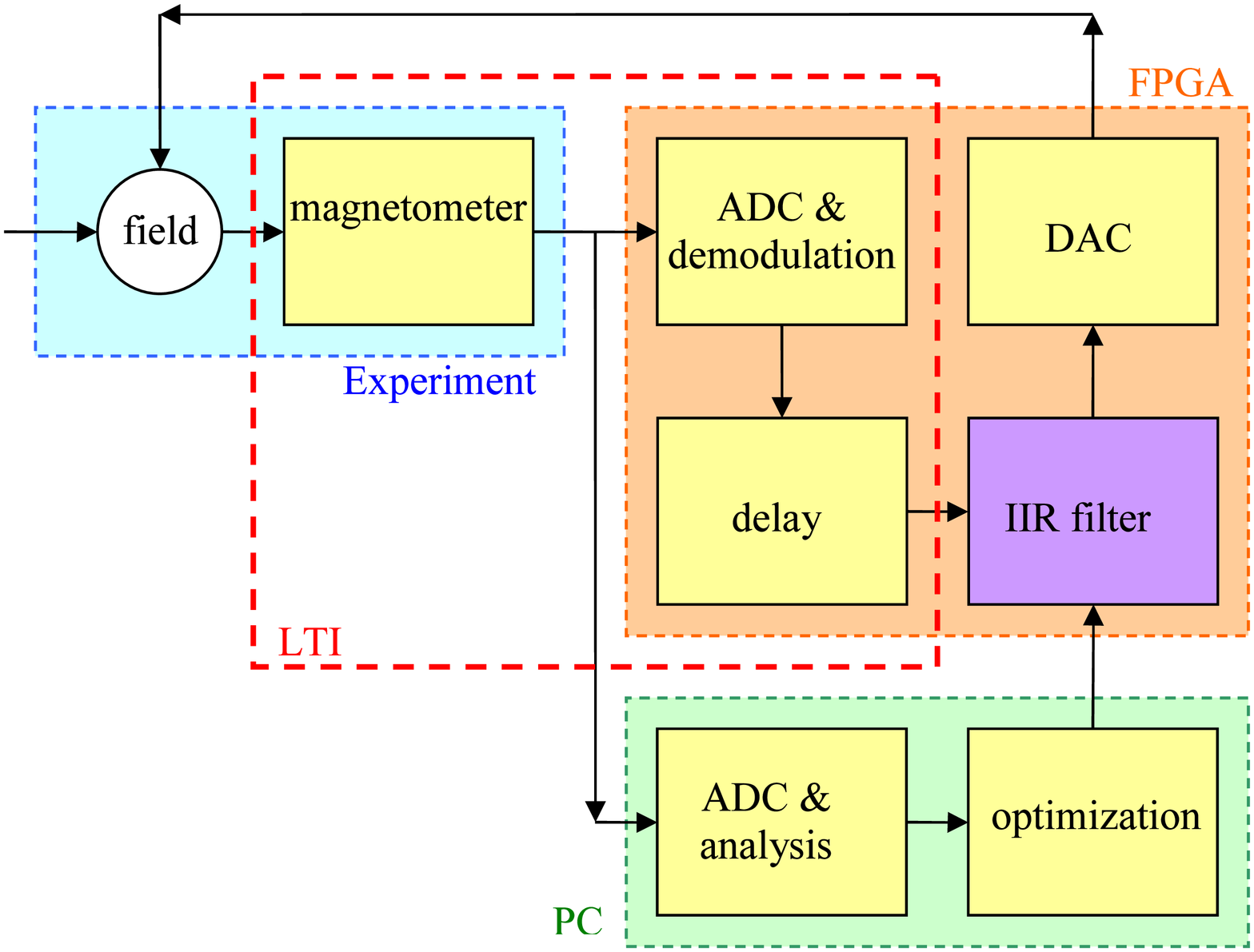}
   \caption{Block diagram of the feedback loop. The system contains three main units: the experiment module -- the magnetometer and the field control (current amplifier and coils); the real time module (FPGA); and the computer for on-line data analysis and optimization. The elements in the dashed box constitute a linear time invariant (LTI) system. A linear control approach for field stabilization through a feedback loop is implemented in the FPGA. An optimization procedure running on the PC determines the IIR filter coefficients for the best noise rejection. }
  \label{fig:loop}
\end{figure}

The block diagram of the  compensation system is presented in Fig.\ref{fig:loop}. The magnetometer followed by the FPGA ADC and demodulator is, to a good approximation, a linear time-invariant (LTI) system with a delayed output, the delay being introduced both at the demodulation/filtering and at the ADC/DAC stages: $\Delta=\Delta_1+\Delta_2$. The loop filter has the two-fold task of compensating both the delay $\Delta$ and the response of the magnetometer, which --to a good approximation--   behaves as a first-order low-pass filter.

\begin{figure}[htbp]
  \centering
   \includegraphics [width= 0.8  \columnwidth] {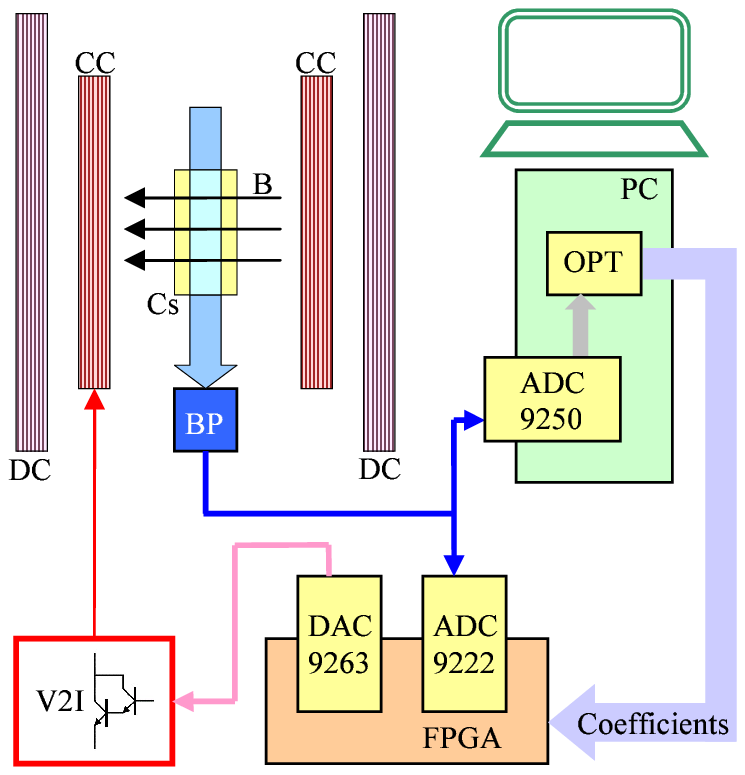}
   \caption{ Sketch of the whole setup. The Cs cell is merged in a field B made of a DC static component, and external disturbances that are counteracted  by a term produced by the compensation coils (CC). The sensor output is digitized both by the FPGA ADC and by a PCI-ADC. The FPGA elaborates the signal to provide a real-time output at the DAC board, which is used to feed the CC through a voltage to current converter (V2I); the PC elaborates finite-time traces to extract the spectral information to evaluate the cost function used in the optimization procedure (OPT). The latter provides loop-filter coefficients to the FPGA and search for their optimal values.}
  \label{fig:wholesetup}
\end{figure}

 The compensation loop is implemented in the FPGA (orange box in Fig.\ref{fig:loop}) where the magnetometric signal is digitized, elaborated and fed back to the experiment in real time. The same signal is also digitized in finite time traces by a 16~bit 1.25~MSa/s PCI-ADC (NI9250) and elaborated in a PC. The trace analysis is performed by the PC and can be used either to extract the magnetometer output, or to evaluate the compensation-loop performance, as necessary for the loop optimization procedure (green box in Fig.\ref{fig:loop}). A general sketch of the whole setup is given in Fig.\ref{fig:wholesetup}, where it is worth pointing out the different roles of the real-time DAC/filtering/ADC operation in the FPGA and the ADC/data analysis/optimization performed in the PC.

In the following subsections we will present the feedback loop performance when (i) the polarimetric signal is manipulated according to the magnetometer frequency response and (ii) in self-adapted regime.

\subsubsection{Loop-filter design based on the magnetometer frequency response}

The network of digital filters in the loop can be categorized in two main types: the finite-impulse-response (FIR), convolution filters characterized by being stable and with a linear phase response (have no analog equivalent);  and infinite-impulse-response (IIR), recursive filters with nonlinear phase response, that are potentially unstable and may have analog equivalent. While a FIR filter is characterized by a single set of coefficients, an IIR filter is characterized by two sets of coefficients (the so called forward and backward coefficients).

The  loop design starts taking into account the OAM sensor response, given by Eq.(\ref{eq:phase}). At resonance ($\omega_M=\omega_L$) and at t=t$_n$ the error signal can be expressed as a FIR output. Namely, $\delta B_n = (\Gamma \varphi_n + \dot \varphi_n)/\gamma $, where  $\dot \varphi$ evaluated at $t=t_n$ in a third-order, finite-difference estimation gives:

\begin{equation}
\dot \varphi_n=\Big(\frac{11}{6} \varphi_n	- 3\varphi_{n-1}	+\frac{3}{2}\varphi_{n-2}	- \frac{1}{3}\varphi_{n-3}\Big)\frac{1}{\Delta_2}.
\label{eq:FIRcoeff}
\end{equation}

This estimation introduces an additional delay $\Delta_3=k\Delta_2/2$ where $k$ is the estimation order ($k=3$ in the case of Eq.(\ref{eq:FIRcoeff})).
The gain factor is  empirically adjusted to achieve the best noise compensation without a significant increase of the noise floor.

\begin{figure}[htbp]
  \centering
  \includegraphics [width= \columnwidth] {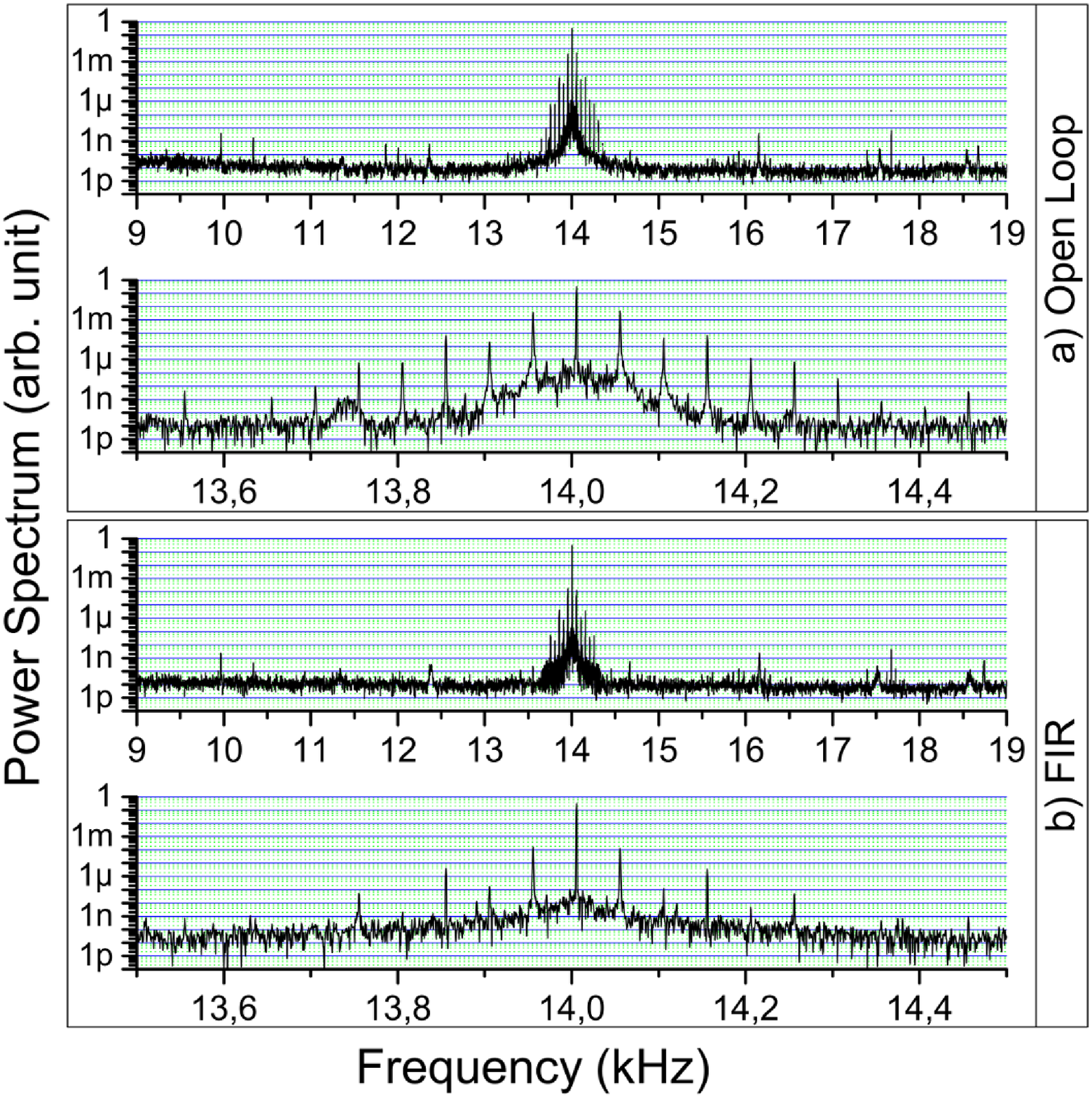}
  \caption{Power spectra, in a wide (upper plot) and a narrow (lower plot) frequency range, of the polarimetric signal of a single magnetometer channel in open loop (subplot a) and in closed loop with an FIR loop filter designed according to the magnetometer frequency response (subplot b).}
  \label{fig:openFIR}
\end{figure}

The open and the closed loop noise patterns, prior to Larmor frequency demodulation, are presented in Fig.\ref{fig:openFIR}. The subplot (a) shows the open loop PSD of the polarimentric signal where, apart from the Larmor frequency peak at about 14~kHz, the low-frequency noise pedestal and discrete noise component at 50~Hz and its harmonics can be seen, in agreement with the demodulated signal presented on Fig.\ref{fig:CMPSD}. Significant improvement of the noise pattern can be observed in subplot (b) when a FIR loop filter is designed according to Eq.(\ref{eq:phase}) with FIR coefficients given by Eq.(\ref{eq:FIRcoeff}). Nevertheless, the pedestal and the discrete noise components still constitute significant noise.

Further improvement of the loop performance can be obtained in a self-adaptive regime - a network of filters is implemented such that its additional degrees of freedom (numerical filters' coefficients) make it possible to adapt the feedback loop filter in order to take into consideration: the detection bandwidth, exact magnetometer response, the actuator (and its supply) response and the introduced delay $\Delta$.

\subsubsection{Optimized feedback loop design}

Delayed error signals constitute a well known problem in control theory\cite{smith_cep_57}. IIR filters provide some tools to address the problem.

An appropriate IIR filter design may indeed enhance the response to the error signal transients, which may help in compensating the effects of the error signal delay. The IIR forward coefficients mainly serve to implement the $\varphi \rightarrow \delta B$ conversion, while the backward coefficients are aimed at compensating the error signal delay. 

The optimization procedure acts on the IIR filter forward and backward coefficients, where the feedback loop performance is evaluated by means of an appositely constructed cost function (CF) described in the following. The evaluation is performed on-line - each set of coefficients works on a new magnetometer data set. Such optimization procedure works well on typical magnetic disturbances.

An ideal output of the self-optimized loop procedure aims to produce a polarimetric signal which is a pure sine-wave at $\omega_M$, i.e. which corresponds to operate in a constant magnetic field. As can be seen from Fig.\ref{fig:CMPSD} the unshielded operation suffers from low frequency noise that is well evident below 100~Hz, and shows the typical $1/f$  increase towards lower frequencies. Some magnetic field components at discrete frequencies are artificially introduced in the range of interest (1~Hz to 100~Hz) to be used in the optimization procedure. The latter maximizes the CF defined as a sum of the ratio between the squared amplitude of the carrier peak at $\omega_M$ and the sum of the squared amplitude of each appositely introduced noise.

In addition, the CF takes into account another aspect of the feedback loops - their instability. In order to avoid loop oscillations or run the feedback oscillation frequency far from the $\omega_M$ the CF is also multiplied by a sigmoid function that turns to zero when the loop resonances exceed a given threshold value at given frequency.

\begin{figure}[htbp]
  \centering
  \includegraphics [angle=0, width= \columnwidth] {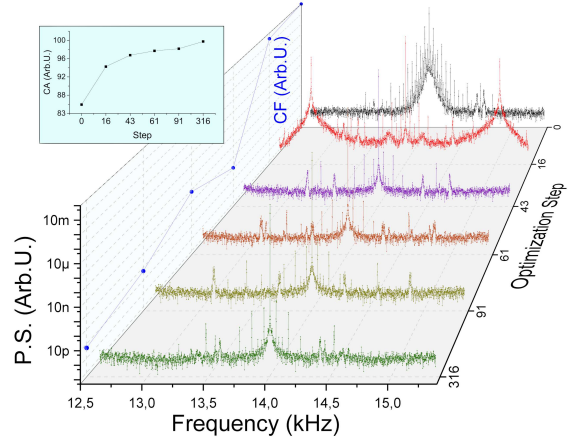}
  \caption{A sample of a few significative steps in the optimization procedure. Starting from the uncompensated case (step 0), a progressive improvement of the polarimetric signal spectra is achieved. The improvement appears in the power spectrum (P.S.) and consists in lower pedestal and peaks around the carrier frequency and in the displacement and reduction of loop resonances, that are obtained while preserving or increasing the carrier amplitude CA (see the inset). The CF reduction is represented in the blue plot on the left vertical plane.  }
  \label{fig:example}
\end{figure}

As an optimization procedure the Nelder-Mead \cite{nelder_cj_65} algorithm is selected, because it does not require the explicit computation of the gradients of the target function with respect to the filter coefficients. The optimization, see the green box in Fig.\ref{fig:loop}, is based on repeated analyses of the spectra of the polarimetric signal. Each spectrum is evaluated from traces lasting about ten seconds, and the whole optimization procedure requires the analysis of several hundred spectra: the overall optimization time lasts typically a few tens of minutes. 
 The Fig.\ref{fig:example} shows a subset of optimization steps to exemplify the progressive improvements of the noise rejection.
At the end of the optimization procedure the artificial disturbances are removed and (provided that the environmental disturbances occur steadily within the spectral range where the optimization has been carried out) the system maintains its rejection efficiency. The actual performance can be tested verifying the efficiency in compensating other sets of artificial disturbances, as well as from the analyses of spectra recorded in the presence of environmental magnetic noise, as those shown later on.

The optimization  of the adaptive loop-filter starts using the $\varphi_n$ (input signal evaluated at $t=t_n$)  estimate to generate a DAC  signal $y_n$ (error signal evaluated at $t=t_n$) in the form:
\begin{equation}
y_n=\sum_{k=0}^{N} b_k \varphi_{n-k}+
    \sum_{j=1}^{M} c_j y_{n-j}
    \label{eq:filtro}
\end{equation}
with the forward coefficients $\{ b_k \}$ designed based on the Eq.(\ref{eq:phase}), and with the backward ones $\{ c_j \}$ enhancing the transient response. According to the time shifting property of the $\mathcal{Z}$-transform\cite{oppenheim_book_09}, generally:
\begin{equation}
\varphi_{n-k}\xleftrightarrow[]{\mathcal{Z}}z^{-k}\Phi[z].
\end{equation}
The transfer function $H[z]$ 
of the filter (Eq.\ref{eq:filtro}) is a rational function given by:
\begin{equation}
H[z]=\frac{Y[z]}{\Phi[z]}=\frac{(b_1+b_2z^{-1}+...+b_Nz^{-N+1})}{(1-c_1z^{-1}-...-c_Mz^{-M})},
 \label{eq:tf}
 \end{equation}
which --in the hypothesis that only simple poles are present-- can be decomposed as a sum of partial fractions:
\begin{equation}
H[z]=\sum_n \Big(A_nz^{-n}+ \frac{B_n}{z^{-1}-q_n}+\frac{C_nz^{-1}}{z^{-1}-p_n} \Big),
\label{eq:pf}
\end{equation}
where $A_n$ are FIR coefficients, $q_n$ and $p_n$ are the  poles in the $z^{-1}$ variable, and $B_n$ and  $p_n C_n$  are the corresponding residues. The decomposition lets speed up the FPGA computation by parallelization, with the optimization algorithm working on the set \{$A_n, B_n, C_n, q_n, p_n$\} instead of \{$b_k, c_j$\}, where typically  $0\le n\le 2$.
 The optimization adapts an IIR filter with a transfer function in the form of Eq.(\ref{eq:pf}). The best noise rejection is obtained applying eight nonzero coefficients among the parameter set.

\section{Results}

\begin{figure}[htbp]
   \centering
  \includegraphics [width= 0.8\columnwidth] {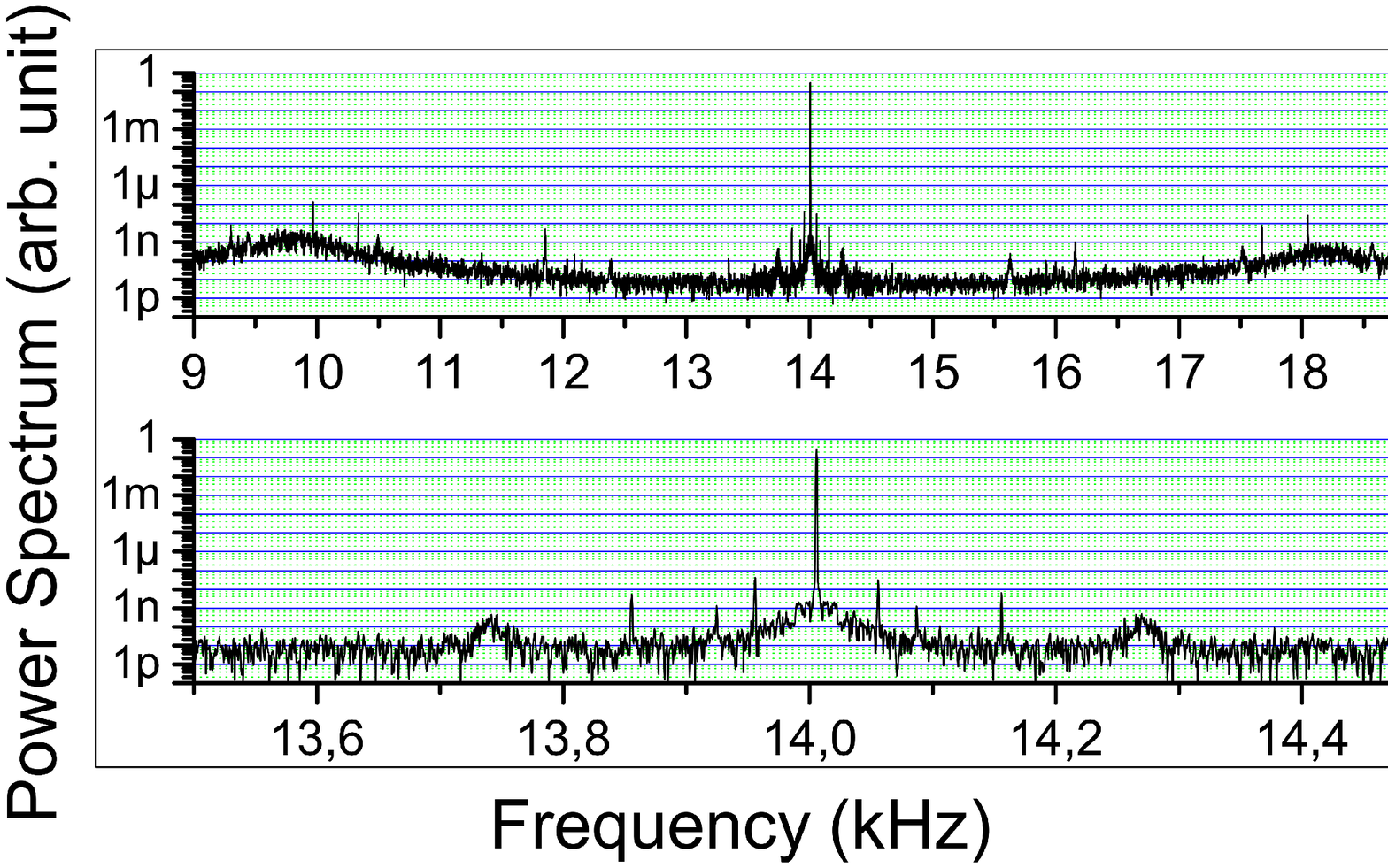}
   \caption{Power spectrum of the polarimetric signal of a single magnetometer channel obtained with an IIR filter after the optimization. Both plots show the same signal in different frequency range. The most prominent feedback loop instability can be seen in the upper plot - around 4~kHz away from the carrier frequency at about 14~kHz. 
  }
  \label{fig:PSD}
\end{figure}

The results of feedback optimization procedure are shown in Fig.\ref{fig:PSD}.
The best CM noise rejection is evaluated taking into account the attenuation of both the discrete peaks and the pedestal, both in the proximity of the carrier frequency, and at frequency displacements well above $\Gamma$. Compared with the subplot (a) of Fig.\ref{fig:openFIR}, the 50~Hz sidebands are attenuated by 15~dB in Fig.\ref{fig:openFIR}.b with the FIR filter and by 50~dB in Fig.\ref{fig:PSD} with the IIR filter. In a narrower frequency range, a pedestal attenuation of 15~dB and 35~dB is achieved, respectively. One hundred Hz away from the peak, the attenuation of the pedestal is less than 10~dB in the case (FIR), while it exceeds 20~dB in case (IIR), attaining the noise floor. Noise increase due to loop instability can be observed in the optimized regime (the noise increase about 4~kHz) but it is possible to keep it at low level and far away enough from the Larmor frequency. 

\begin{figure}[htbp]
   \centering
  \includegraphics [width=\columnwidth] {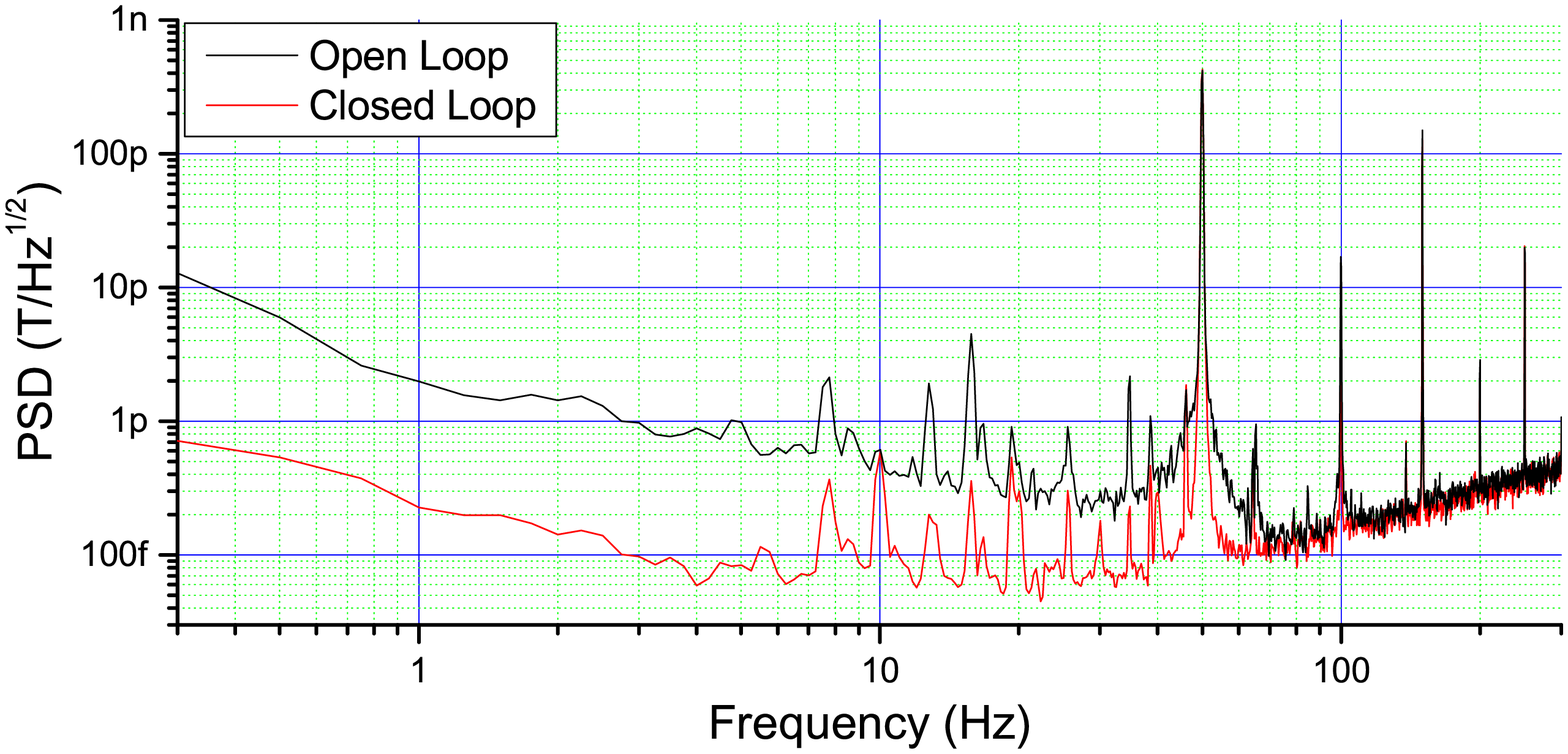}
  \caption{Power spectrum density of the demodulated magnetometric DM signal.  
  }
  \label{fig:noise_dif}
\end{figure}

The compensation efficiency in optimized regime has a broadband nature and is not markedly dependent on the frequency of the appositely introduced disturbances for feedback filter coefficients optimization. This also guarantees a robust behavior in response to any environmental disturbance occurring occasionally within the spectral range where the optimization has been performed. 

\section{Application}
The considered application of the whole setup (multichannel magnetometer and active magnetic field compensation system) is the measurement of NMR signals in ULF range. This kind of experiment is performed in a condition where the magnetized nuclei are placed in the proximity of one of the channel, in such a way that their precessing magnetization generates maximum differential signal. Operating at the $\mu$T level,  NMR spectral features lie in the 1-100~Hz range. Several interesting peculiarities make this kind of NMR spectroscopy of interest: it is technically easy to produce low magnetic field with excellent spatial homogeneity, which dramatically reduces the instrumental line broadening\cite{tayler_rsi_17}; different nuclear species can be detected simultaneously\cite{bevilacqua_jpcl_17}; the energy of the interaction between different nuclei in molecules becomes comparable to the interaction of nuclei with the static field. On the other hand, operating at such low fields, makes the external disturbance a severe problem, both because of the noise introduced at the detection stage has a spectrum superimposed on that of the nuclear signal, and because if the nuclear precession occurs in time dependent field this hinders the application of averaging techniques for the S/N enhancement.

The final scope of the presented methods is to improve the signal-to-noise ratio in DM mode measurements by an effective CM disturbance attenuation. In Fig.\ref{fig:noise_dif} the DM spectra  registered in open loop and in closed-optimized-loop conditions are shown: an improvement by more than a factor 10 in the disturbance rejection is observed at the lower frequencies (300~mHz to 5~Hz), which then progressively reduces at higher frequencies, up to about 100~Hz, where the it attains the intrinsic noise level.

\section{Conclusion}
The proposed methodology offers a powerful tool for designing and optimizing feedback loops, suited to achieve robust stabilization apparatuses when dealing with poorly known system responses and/or when high  noise-rejection efficiency is required. The approach has been studied to be applied in high sensitive multichannel magnetometry, however it can find application in any differential system that can take advantage of a feedback-based counteraction of the CM noise. Our findings demonstrate the effectiveness of the  approach for improving the CM magnetic noise rejection over a broad spectral range, which results in a noticeable reduction of the noise level in DM measurements.

\acknowledgments
The authors thank Prof. Marco Maggini for clarifying and fruitful discussions.

\bibliographystyle{ieeetr}
\bibliography{refs}

\end{document}